\documentclass[reprint,
superscriptaddress,
aps,
pra,
showpacs,
]{revtex4-2}

\usepackage{color}
\usepackage{graphicx}
\usepackage{dcolumn}
\usepackage{bm}
\usepackage{diagbox}

\DeclareUnicodeCharacter{2212}{-}

\begin{document}
\preprint{APS/123-QED}

\title{Random Sampling Neural Network for Quantum Many-Body Problems}

\author{Chen-Yu Liu}
\affiliation{National Center for Theoretical Sciences, Hsinchu 30013, Taiwan}

\author{Daw-Wei Wang}
\affiliation{National Center for Theoretical Sciences, Hsinchu 30013, Taiwan}
\affiliation{Department of Physics, National Tsing Hua University, Hsinchu 30013, Taiwan}
\affiliation{Center for Quantum Technology, National Tsing Hua University, Hsinchu 30013, Taiwan}                   

\begin{abstract}
The eigenvalue problem of quantum many-body systems is a fundamental and challenging subject in condensed matter physics, since the dimension of the Hilbert space (and hence the required  computational memory and time) grows exponentially as the system size increases. A few numerical methods have been developed for some specific systems, but may not be applicable in others. Here we propose a general numerical method, Random Sampling Neural Networks (RSNN), to utilize the pattern recognition technique for the random sampling matrix elements of an interacting many-body system via a self-supervised learning approach. Several exactly solvable 1D models, including Ising model with transverse field, Fermi-Hubbard model, and spin-$1/2$ $XXZ$ model, are used to test the applicability of RSNN. Pretty high accuracy of energy spectrum, magnetization and critical exponents etc. can be obtained within the strongly correlated regime or near the quantum phase transition point, even the corresponding RSNN models are trained in the weakly interacting regime. The required computation time scales linearly to the system size. Our results demonstrate that it is possible to combine the existing numerical methods for the training process and RSNN to  explore quantum many-body problems in a much wider parameter regime, even for  strongly correlated systems.
\end{abstract}

\maketitle

\section{Introduction}

It has been a long-standing challenge in condensed matter physics that the eigenvalue problem of a many-body system is in general not accessible because of the exponentially huge Hilbert space of the associated quantum many-body Hamiltonian. Only very few simple models in low dimensional space acn be solved exactly due to their higher order symmetries \cite{pierre69, cnyang1, bm}. As a result, various analytical and numerical methods are developed for certain specific systems, including perturbation theory \cite{pt, pt2},
renormalization group \cite{rg1, rg2, rg3}, bosonization \cite{bz1, bz2}, Quantum Monte Carlo \cite{Raimundo03, qmc1, qmc2}, Density Matrix Renormalization Group \cite{dmrg1, dmrg2} and tensor networks \cite{Orus17, tn1} etc. Within these analytic or numerical methods, there are also many exquisite techniques developed for solving some specific many-body problems in some parameter regimes. Furthermore, in recent rapidly growing development of machine learning approaches \cite{ds}, certain unsupervised learning methods, such as Neural Network Quantum State (NQS), are found to have better results than ordinary variational methods in the calculation of the ground state and excited state energies \cite{Gius17, Kenny18} through its undetermined parameters of restricted Boltzmann Machine. However, in these various approaches, each data point is calculated independently according to the associate system parameters, and therefore may cost a lot of computational resources to get a complete phase diagram. 

From the data-driven machine learning point of view, on the other hand, this problem could be investigated from different perspectives. Instead of unsupervised learning approaches, which are similar to variational methods, one could also train a model based on existing results in a well-known parameter regime and then apply this model to other regime. Most applications along this line are the classification and/or feature extraction of various phase transitions through the supervised learning approach \cite{Broecker17, Kelv17, Ming19, mlp}. 
The basic concept is to train a model to learn the identities (i.e. labels) of different phases in some parameter regimes, and apply it to define the phase boundary in the middle regime. 
Such a numerical approach is possible because a horizontal relationship between features in different parameter regimes could be learned (more precisely, fitted by a complicated function through machine learning) during the training process. However, since the input features are usually from experimental data or other physical quantities, and the labels of these phases are in fact artificial labelling for the purpose of classification (say $=(0,1)$ for phase A and $=(1,0)$ for phase B), such kind of pattern recognition scenario could not provide sufficient information for the understanding of a many-body system. After all, the ``nature" of these phases should be associated with the relationship between these physical quantities, while such a relationship is in general too complicated to be calculated in most systems.

Combining the advantages of the two approaches mentioned above, in this paper, we propose a self-supervised machine learning method, Random Sampling Neural Network (RSNN), to study a general many-body problem. Motivated by pattern recognition and self-supervised methods in computer vision \cite{pra1, pra2, sf1}, we treat the many-body Hamiltonian as a huge 2D "system image". Random sampling matrix elements (i.e. "patches") are collected from this "system image" as the input features to train a Convolutional Neural Network (CNN) in the training regime. The labels of these features are physical quantities obtained by the system Hamiltonian at the same parameter, reflecting the spirit of a self-supervised learning process. We show that the accuracy of such simulation in the test regime can be systematically improved by increasing the amounts of training data, while the computation time just scales linearly to the system size, no matter what kinds of systems or physical quantities to study. We use several 1D exactly solvable models, including Ising model with a transverse field, Fermi-Hubbard model, and spin-$1/2$ $XXZ$ model, to demonstrate the applicability of RSNN in the strongly correlated regime, if only the model is properly trained by known results (obtained by other numerical methods) in the weakly interacting regime. Our results show that RSNN can be an efficient data-driven method and hence also a complementary approach to the existing analytic/numerical methods for the study of quantum many-body problems. 

In the rest of this paper, we will first introduce the basic concept and hypothesis of RSNN in Sec. \ref{sec:concept}, and then use 1D Ising Model with a transverse field as the first example of RSNN in Sec. \ref{sec:IMTF}. In Sec. \ref{sec:accuracy and computation time}, We then use 1D IMTF to systematically investigate how the accuracy and computation time changes for different hyper-parameters of RSNN. Similar results should be also expected for other physics models. We then apply RSNN to predict the whole energy spectrum of 1D Fermi-Hubbard Model in the strongly correlated regime in Sec. \ref{sec:FHM}. In Sec. \ref{sec:XXZ}, we further apply RSNN to predict the quantum phase transition point of the 1D $XXZ$ model and investigate its quantum critical exponent. In Sec. \ref{sec:summary}, we summarize our results and then provide a Github code for the application of RSNN in 1D Ising Model with a Transverse Field. Further details of the RSNN models are shown in the Appendix.

\begin{figure}
\includegraphics[scale=0.21]{"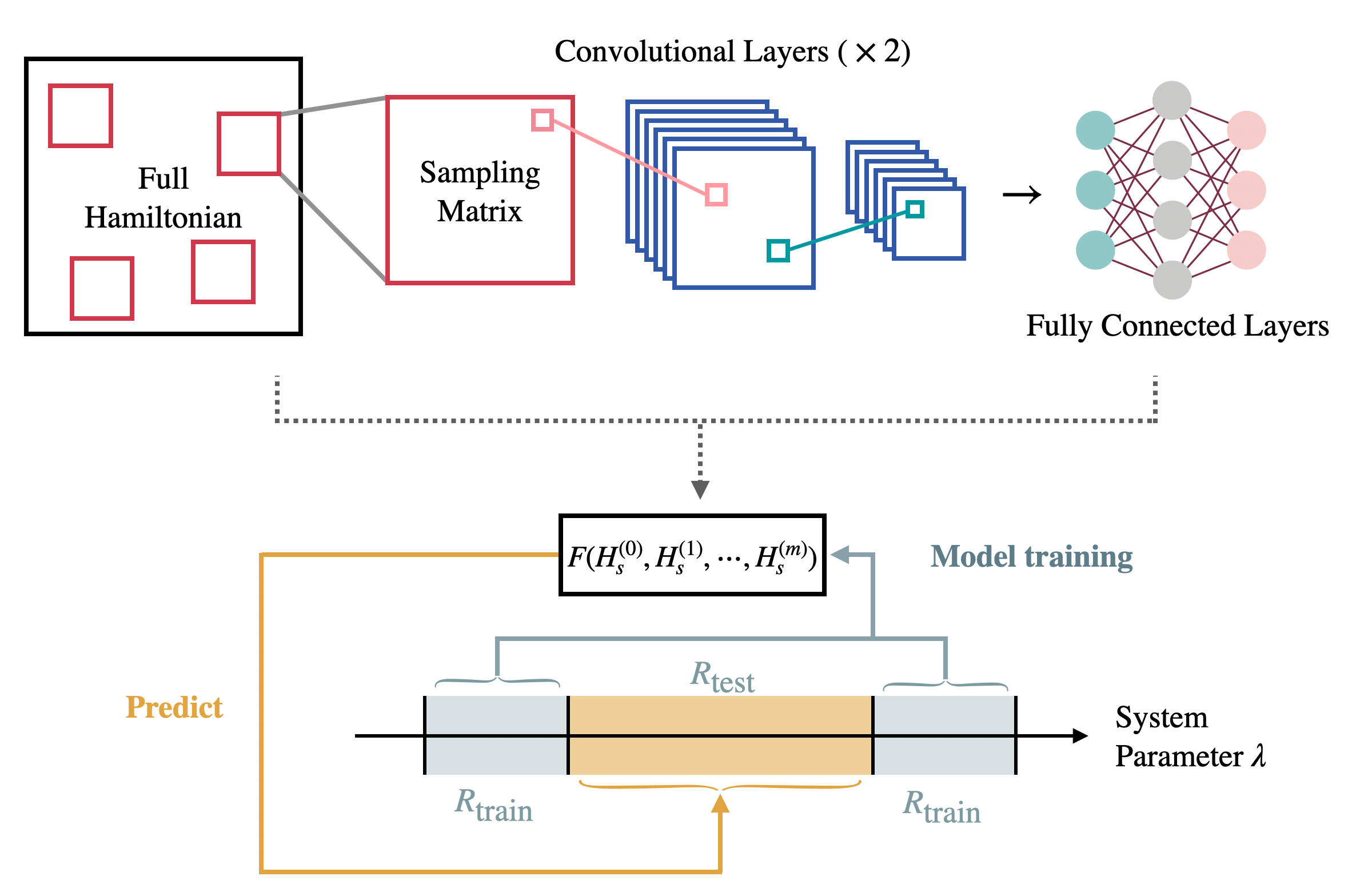"}
\caption{
A typical flowchart of a RSNN model. The input data are $M\times M$ matrices, $\textbf{H}^{(m)}_S$, randomly sampled from the original full system Hamiltonian (see the text). After a standard self-supervised learning by a CNN model with known results in the training regime ($R_\textrm{train}$), the obtained RSNN model can be used to predict physical results in the test regime ($R_\textrm{test}$). In most physical problems, the training and test regimes can be defined by a continuous parameter, $\lambda$. 
}
\label{fig:method}
\end{figure}
\section{Basic Concept and Hypothesis}
\label{sec:concept}

The concept of RSNN is motivated by machine learning methods developed for the patter recognition: The system Hamiltonian ($\hat{\cal H}$) can be treated as a 2D "system image", after represented by a hermitian matrix ($\textbf{H}$) with matrix elements, $H_{\mu\nu}=\left<\phi_\mu\right|\hat{\cal H}\left|\phi_\nu\right>$. Here $\{\left|\phi_\nu\right>\}$ is a complete and orthonormal basis. Each $H_{\mu\nu}$ is like a "two-color pixel" with its real and imaginary parts. As a result, the process to calculate any physical quantities (for example, its eigenenergies, $E_n$, or any expectation value of the ground state) is thus equivalent to derive the functional relationship between the matrix elements and these quantities, i.e. $F(\textbf{H}[\lambda])=F(\{H_{\mu\nu}(\lambda)\})=\{E_n(\lambda)\}$, where $\lambda$ stands for a system parameter in control (for example, external magnetic field or coupling strength). From the machine learning point of view, interestingly, solving such a functional relationship ($F$) is thus equivalent to training a neural network model ($F_\textrm{NN}$) to simulate this complicated function, i.e.
\begin{eqnarray}
F_\textrm{NN}\left(\textbf{H}[\lambda]\right)\Rightarrow F(\textbf{H}[\lambda])=\{E_n(\lambda)\}
\end{eqnarray}
According to the universal approximation theorem \cite{uat1}, the difference between the approximate function ($F_\textrm{NN}$) and the true function ($F$) can be infinitesimal if the number of artificial neurons (and hence the fitting parameters) and training data used for $F_\textrm{NN}$ are large enough. 

In a general many-body system, however, the dimension of such a "system image" (i.e. the matrix representation of the system Hamiltonian) grows exponentially as the system size increases and therefore cannot be simulated efficiently. To overcome this problem, here we propose a random sampling method, where a large image could still be recognized if only patches of this image are used during the training process. More precisely, we first randomly select $M$ basis vectors from the full many-body basis to construct an $M\times M$ sampling matrix, $\textbf{H}_S^{(m)}$, and repeat this sampling for $N_S$ times ($m=1,\cdots N_S$). The selected basis for each time can be different from each other. These random sampling matrices therefore form a collection of "patches" of the original ``system image", and hence contain partial information of the full many-body system.

We could then combine these random sampling data and the concepts of a self-supervised learning model, which is trained by the internal properties of a system rather than external labels \cite{exl}, and propose the following \textit{Random Sampling Hypothesis}: The "patches" of the full "system image", extracted from random sampling basis, can be used as the input features of a neural network model, so that, in a given training regime ($\lambda\in R_\textrm{train}$), the obtained random sampling function, $F_\textrm{RSNN}$, can simulate the target physical quantities via a self-supervised learning process within a small deviation: 
\begin{eqnarray}
\left| F_\textrm{RSNN}\left(\{\textbf{H}_{S}^{(m)}[\lambda],\textbf{b}_S^{(m)}[\lambda]\}\right)- \{E_n(\lambda)\}\right|<\epsilon.
\label{eq:hypothesis}
\end{eqnarray}
Here the upper bound of their difference, $\epsilon$, can be reduced if only the amounts of artificial neurons and/or the training data are increased. Its application in the test regime ($\lambda\in R_\textrm{test}$) can therefore also provide reliable estimates if $R_\textrm{test}$ is not too far from $R_\textrm{train}$.

Before applying this hypothesis to a realistic physical problem, we have to emphasize that the collection of sampling basis ($\{b_S^{(m)}[\lambda]\}$) can be different for each time, so that the neural network could be enforced to simulate the eigenvalue-solving problem based on the partial information of the selected basis. Similar scenario can be also applied to the calculation of other physical quantities, such as magnetization or spectral function etc.  In the rest of this paper, we will provide numerical calculation to support this hypothesis and investigate its application in different many-body problems.  

\begin{figure}
\includegraphics[scale=0.33]{"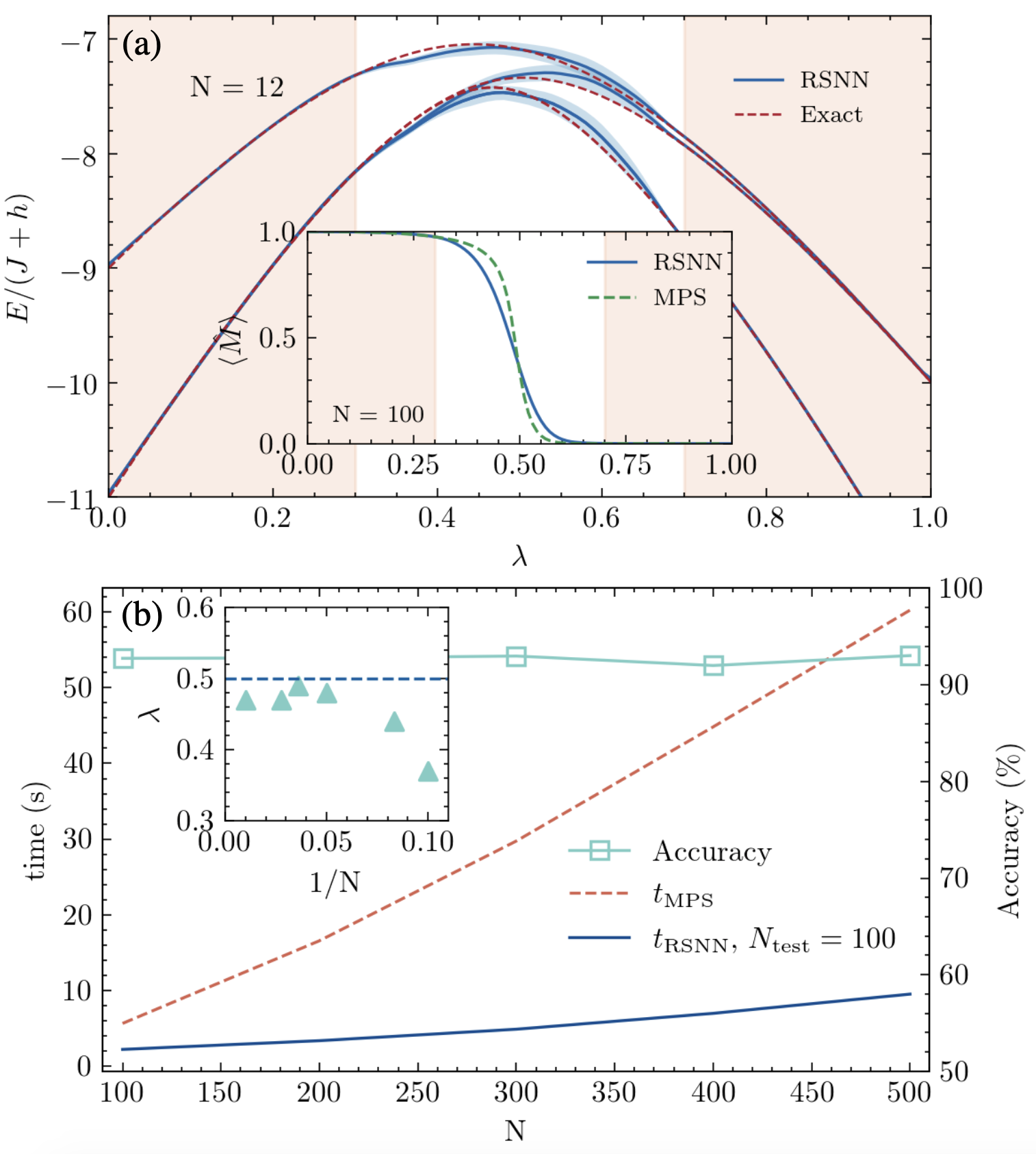"}
\caption{
(a) Prediction result of the lowest three eigenenergies of 1D IMTF by RSNN (solid lines) for $N = 12$. The training regime, $R_\textrm{train}=(0,0.3)\cup(0.7,1.0)$, is shown by the colored background, and the test regime, $R_\textrm{test}=(0.3,0.7)$ is by the white background. The blue gray area near the RSNN result indicates the uncertainty, resulting from five independent calculations. The results obtained by the exact diagonalization (dashed lines) are shown together for comparison. The inset shows the predicted magnetization (solid line) for $N=30$, compared to the results obtained by MPS (dashed line). Here we use $N_{train}=10000$, $N_S=200$, and $M = 10$ for the training process (see the text). Note that we have scaled the energy in unit of $J+h$ in order provide a more balanced expression of eigenstate energies in the training regime. (b) shows the average computation time (solid line) to generate a test data by RSNN for $N_{test}=100$ data points (see the text). The computation time by MPS (dashed lines) as a function of system sizes are also shown together for comparison. Accuracy of the magnetization of RSNN method are also shown together by open squares. The inset shows the finite size scaling of the phase transition point, which is defined when the separation between the lowest two eigenenergies is larger than their uncertainties. Horizontal dashed line is the quantum critical point ($\lambda_c=0.5$) in the thermodynamic limit. Other parameters are the same as (a).}

\label{fig:1D_IMTF_main}
\end{figure}
\section{Ground state and Magnetization of 1D 
Ising Model with a Transverse Field:}
\label{sec:IMTF}

In order to demonstrate the applicability of RSNN, we first take 1D spin-$1/2$ Ising Model with a Transverse Field (IMTF) \cite{pierre69} as an example for systematic studies. The system Hamiltonian is known to be:
\begin{eqnarray}
\hat{\cal H}_\textrm{IMTF} = -J \sum_i^N \sigma_i^z \hat{\sigma}_{i+1}^z - h\sum_{i}^N \hat{\sigma}_i^x,
\end{eqnarray}
where $\hat{\sigma}_{x,y,z}$ are Pauli matrices, $J>0$ is the spin coupling between the nearest neighboring site, $h$ is the transverse field strength, and $N$ is the total number of spins. 1D IMTF with the periodic boundary condition is known exactly solvable though the Jordan-Wigner transformation \cite{jwt}, and therefore could be a good example to test the application of our RSNN. In the thermodynamic limit, the ground state is a double degenerate ferromagnetic phase as $h<J$, and becomes a non-degenerate paramagnetic phase as $h>J$. We could then define $\lambda\equiv h/(h+J)\in (0,1)$ as a dimensionless system parameter (see Fig. \ref{fig:method}) to measure such a phase transition.

In Fig. \ref{fig:1D_IMTF_main}(a) we show the results of RSNN for the three lowest eigenenergies of $N=12$. (Note that the energy is scaled by $J+h$ in order to have a better expression of the energy in the both sides of training regime. We select $N_\textrm{train}=10000$ values of $\lambda$s from the training regime (colored background), and generate $N_s=200$ sampling matrices (with a dimension $M=10$) for each $\lambda$ as the training data, labeled by the exact eigenstate energies. Since the ground state is pretty known in the regime as $\lambda\to 0$ and $\lambda\to 1$, we choose the training regime in both sides and predict results in the middle (test) regime, $R_\textrm{test}= (0.3,0.7)$, where a first order  quantum phase transition is expected to appear around $\lambda=0.5$ in the thermodynamic limit. Comparing the results of RSNN to the exact results in the test regime, we fine the accuracy, $\textrm{Acc}\equiv 1-Ave\left[\sum_{i=0}^2|E_i^\textrm{RSNN}-E_i^\textrm{ED}|/E_i^\textrm{ED}\right]= 99.43 \pm 0.19 \%$. Here $E^\textrm{RSNN/ED}_i$ is the $i$-th eigenvalue obtained by RSNN and exact solution respectively, and $Ave[\cdots]$ is the average taken in the whole test regime for five independent calculations. The inset of Fig. \ref{fig:1D_IMTF_main}(a) shows the predicted magnetization by RSNN for $N = 30$, where the training values of magnetization are calculated by Matrix Product State (MPS) method~\cite{mps1, mps2}. As we could see that the eigenstate energies and magnetization predicted by RSNN are very close to the exact results. Below we will use this model to investigate the accuracy and efficiency of RSNN in various conditions. Details of these model parameters are shown in Appendix \ref{sec:model}.

In Fig. \ref{fig:1D_IMTF_main}(b), we show the average computation time of each data for RSNN ($t_\textrm{RSNN}$) as a function of the system size, $N$. Computation time by MPS ($t_\textrm{MPS}$) is also shown together for comparison. Here $t_\textrm{RSNN}$ is calculated by adding all the generation time of random sampling matrices (for both the training data and the test data) as well as the training time of the RSNN model, and then divided by the total number of test data ($N_\textrm{test}=100$). For comparison, we also show the calculation time of MPS (dashed line) in the same plot. We find that different from the exponentially growing time by exact diagonalization (not shown here) or the long computation time of MPS in the large $N$ limit, $t_\textrm{RSNN}$ grows much more slowly and linearly in the large $N$ lime, while the accuracy of output is still above 99\% for the eigenvalues (not shown here) and above $92\%$ for the magnetization even for $N>100$. Note that, since the required computation time for the feature generation and model training of RSNN is almost the same no matter how many test data are generated, RSNN could be a very efficient method to generate much more test data in the whole parameter regime within a reasonably good accuracy, even for a large system size. 

The slow growing rate of $t_\textrm{RSNN}$ as a function of the system size $N$ can be understood as follows: the preparation time of each data sampling ($t_{S}$) depends on the system size linearly for the calculation of matrix elements, while the training time ($t_\textrm{train}$) depends on the model parameters as well as training scheme only. These two time scales determine the time for data preparation, but are not sensitive to the system size. That is why RSNN could be more efficient than other numerical methods especially for a larger system size. In the inset of Fig. \ref{fig:1D_IMTF_main}(b), we also show the finite size scaling of the phase transition point, which is defined when the difference of the lowest two eigenenergies is larger than their uncertainty. We could find that the precise determination of phase transition point is also possible due to the generation of a large amount of data within reasonable accuracy.

\begin{figure}
\includegraphics[scale=0.37]{"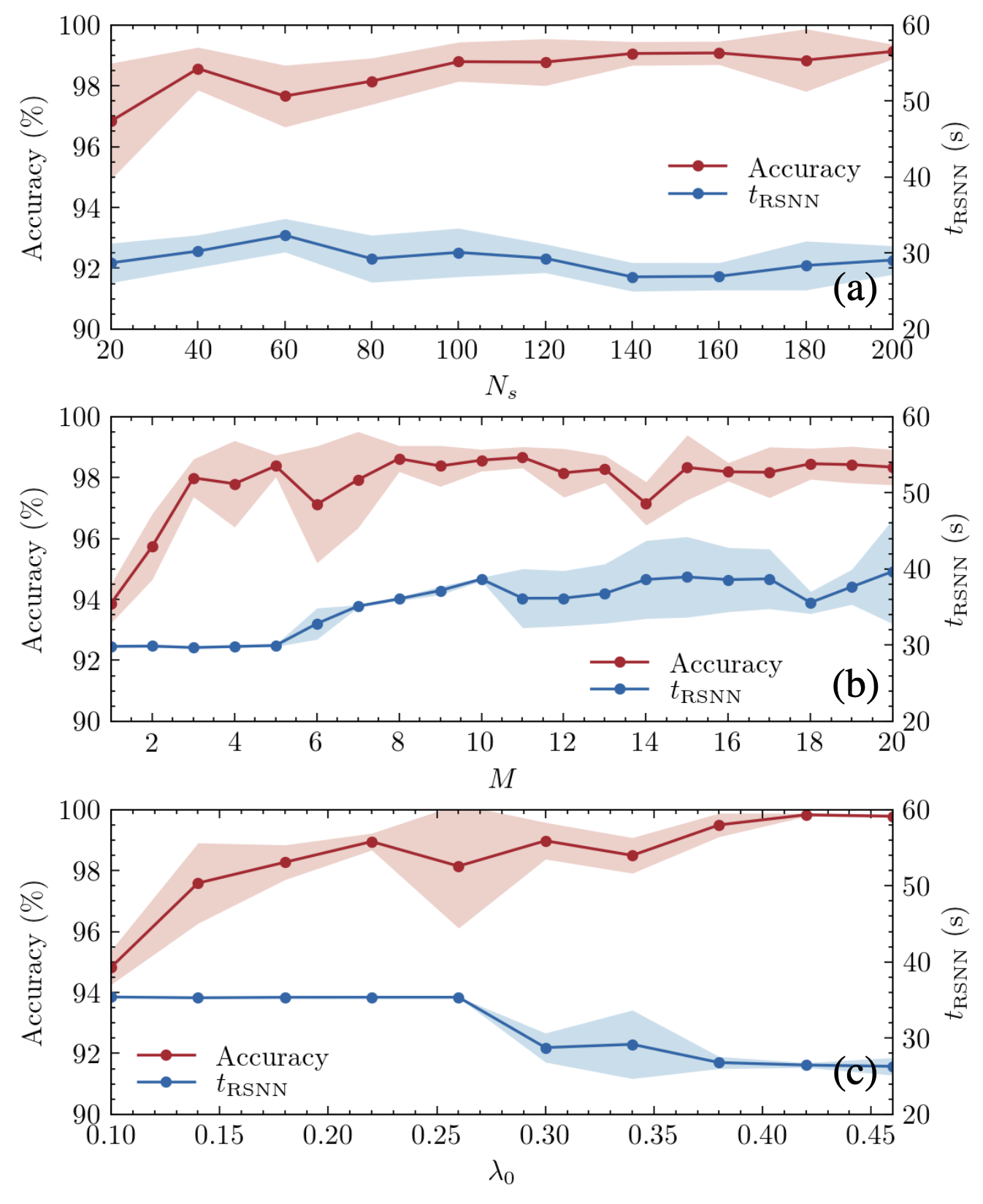"}
\caption{
(a) Accuracy and average computation time of RSNN ($t_{RSNN}$) for the lowest three eigenenergies of 1D IMTF as a function of the number of sampling ($N_s$) at each $\lambda$. Here $N = 28$, $N_S = 200$ and $M = 10$. (b) shows the same calculation but for different size of sampling matrix, $M$, for $N_S=50$ and $N = 28$. The average computation time stop increasing due to the early stop mechanism when the training loss is saturated. (c) shows the same calculation for different value of training regime, $\lambda_0$ (see the text). The total number of $\lambda$s for different training regimes are still the same ($N_{train}=10000$). Here $N_S=200$, and $M=10$ and $N=28$.
}
\label{fig:1D_IMTF_Ns_M}
\end{figure}

\section{Accuracy and Computation Time for Different Hyper-Parameters}
\label{sec:accuracy and computation time}

From the basic calculation shown above for 1D IMTF, we have demonstrate the possibility to apply RSNN for the study of a quantum many-body system. However, we have to emphasize that the results obtained above is non-trivial, especially for a large system size, because the the Hilbert space grows exponentially as $N$ increases, and hence only exponentially small fraction of matrix elements are included in the RSNN. Therefore, the success of RSNN here results from the fact that the obtained simulation function, $F_\textrm{RSNN}$, does capture how the eigenvalues and/or other physical parameters change as a function of the system parameter, $\lambda$, through a small portion of matrix elements. In order to demonstrate this, below we systematically investigate how the accuracy and efficiency of RSNN can change by tuning the hyper-parameters of RSNN during the training process.

In Fig. \ref{fig:1D_IMTF_Ns_M}(a), we show how the average accuracy of eigenstate energies increases as the number of sampling matrices ($N_S$) increases. This reflects the fact that including more sampling matrices shall enhance the accuracy as the nature of neural networks. This also implies that such a high accuracy of prediction in the test regime could not be obtained faithfully if one just fits the eigenstate energy curves without knowing the matrix elements (i.e. $N_S\to 0$). This result therefore, demonstrate the validity of our \textit{random sampling hypothesis} as described in Eq. (\ref{eq:hypothesis}). Furthermore, we find that the average computation time ($t_\textrm{RSNN}$) does not increase but eventually becomes saturated for large $N_S$, because we have applied the early stop method during the training process to avoid over-fitting. In (b), we show the same calculation as a function of sampling matrix dimension, $M$, with $N_S=200$ being fixed. We also find that the computation time grows up significantly in the small $M$ regime, but becomes saturated as $M>10$ due to early stop mechanism. Note that, comparing to the calculation for larger values of $N_S$ in (b), it requires much more computational memory in RSNN when the dimension of sampling matrices ($M$) increases, since the input features (matrix elements) scale as $M^2$. Therefore, here we just show the calculated results upto $M=20$ and expect the accuracy could grow further for a larger value of $M$. 

Finally, in Fig. \ref{fig:1D_IMTF_Ns_M}(c), we show how the accuracy and average computation time changes as a function of $\lambda_0$, which measures the relative size of training regimes by $R_\textrm{train}\equiv (0,\lambda_0)\cup (1-\lambda_0,1)$, see Fig. \ref{fig:1D_IMTF_main}(a). Here we have fixed the total number of the training data ($N_\textrm{train}$) in the training regime with $N_S=200$ and $M=10$ for all different values of $\lambda_0$. As expected, the calculated results shows that the overall accuracy of the RSNN prediction increases monotonically as a function of $\lambda_0$, and reaches 100\% when $\lambda_0\to 0.5$, because the test regime is so close to the training regime. On the other hand, $t_\textrm{RSNN}$ still keeps almost a constant since the total number of training data is the same. 

We evaluate our models on Google Colaboratory cloud computing platform, which specifies two cores Intel(R) Xeon(R) CPU @ 2.20GHz and NVIDIA Tesla P100 GPU.

\begin{figure}
\includegraphics[scale=0.27]{"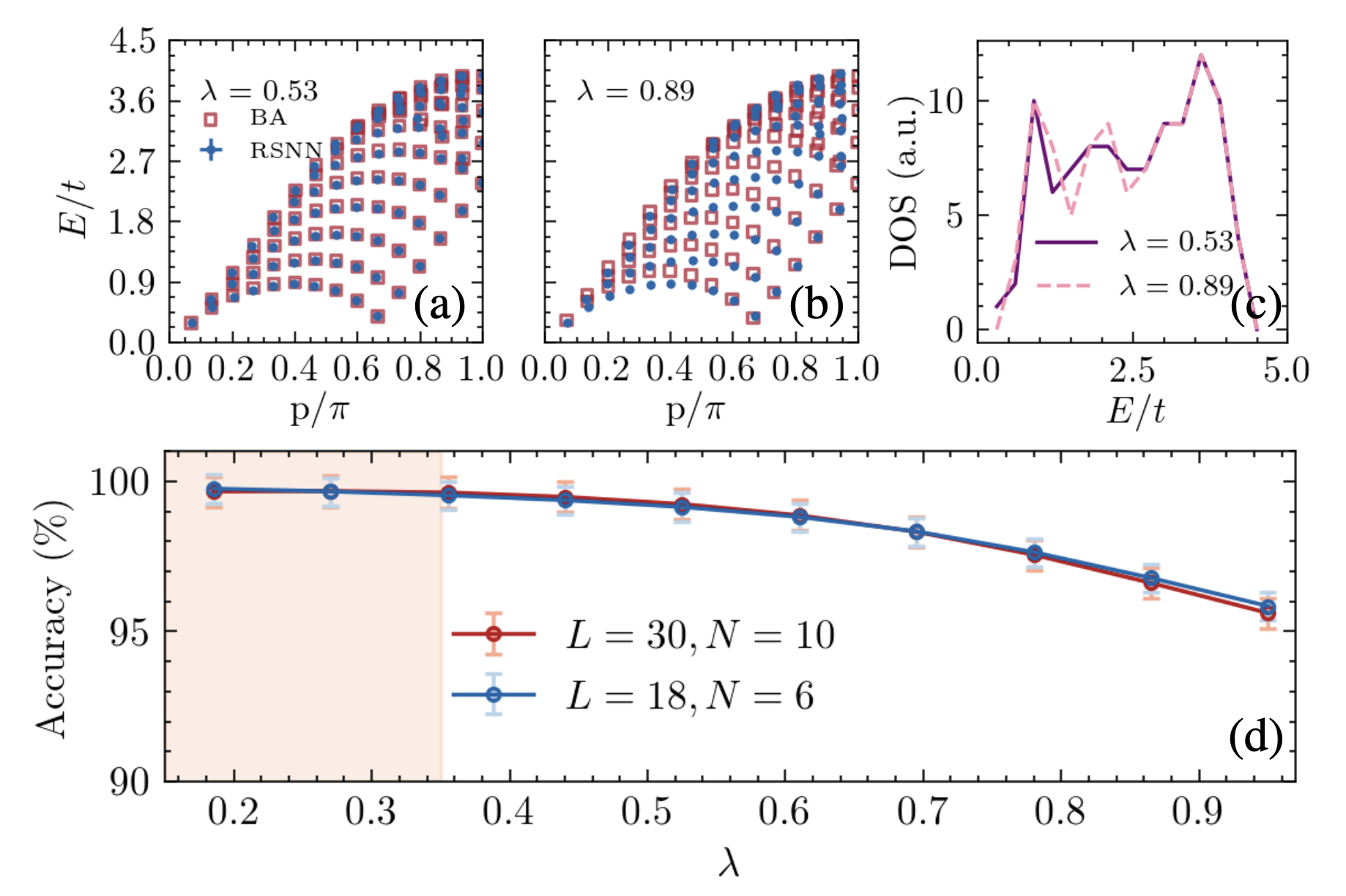"}
\caption{
(a) and (b) are predicted holon excitation spectrum (blue dots) of 1D FHM by RSNN for $\lambda\equiv U/(t+U)= 0.53$ and 0.89 respectively. The training regime is $R_\textrm{train}=(0.1,0.35)$ with $L=30$, $N=10$ and $N_{\downarrow}=5$. Exact results calculated by Bethe-ansatz (BA, red open squares) are also shown together for comparison. (c) shows the associate density of states (DOS) obtained by RSNN. (d) shows the accuracy of the energy spectrum with an uncertainty by RSNN in the test regime, $R_\textrm{test}=(0.35,1.0)$ (white background). Results for two different system sizes are shown together for comparison. 
}
\label{fig:FH_L18N6_DL_pred}
\end{figure}
\section{Energy Spectrum in the strongly interacting regime of 1D Fermi-Hubbard model:}
\label{sec:FHM}

The 1D IMTF discussed above is a good example for the application of RSNN, but it is not a strongly correlated system, because its Hamiltonian can still be mapped to a free fermion model via the Jordon-Wigner transformation \cite{jwt} and hence the eigenstates are still product states without correlation. In order to investigate the application of RSNN in the strongly correlated regime, here we consider 1D Fermi-Hubbard model (1D FHM) with the following system Hamiltonian:
\begin{eqnarray}
\hat{\cal H}_\textrm{FH} = -t\sum_{i,s}\left(\hat{c}_{i,s}^\dagger \hat{c}_{i+1,s}+h.c.\right)+U\sum_{i}\hat{n}_{i,\uparrow}\hat{n}_{i, \downarrow},
\label{eq:FH}
\end{eqnarray}
where $\hat{c}_{i,s}$ and $\hat{n}_{i,s}$ are the fermion field operator and the number operator at site $i$ and of spin $s=\uparrow/ \downarrow=\pm$. $t$ and $U$ are the hopping energy and the on-site repulsion respectively.

It is well-known that in the weakly interacting limit ($U/t\to 0$), 1D FHM can be well-described by a Luttinger liquid \cite{fhmll}, where all elementary excitations are bosonic and collective modes, which can be separated into the spin and charge sectors (i.e. spin-charge separation). When considering the backward scattering for a large momentum transfer as well as the Umklapp scattering in the presence of a periodic lattice, the spin/charge excitations become gapless at the momentum $p=2k_F$ and $4k_F$ respectively even for an infinitesimal $U>0$.  In the strongly interacting limit ($U/t\to\infty$), on the other hand, states with different particle distributions are almost degenerate with either zero or one particle per-site. The system then becomes equivalent to the $t-J$ model \cite{fhmtjm} with an anti-ferromagnetic spin-exchange coupling through the second order perturbation of $t$ (i.e. $J\propto t^2/U$). 

Below we will use 1D FHM as an example to test RSNN in the strongly correlated regime, after the model is trained in the weakly interacting regime. More precisely, we train a RSNN model using the exact results of momentum-energy dispersion, obtained by the Bethe-ansatz method (BA)\cite{fhba1, fhba2}. The non-linear coupled algebraic equations within BA method is described by
\begin{eqnarray}
&&e^{ik_j L} = \Pi_{\alpha = 1}^{N_\downarrow} \frac{\sin k_j - \lambda_\alpha + i U/4}{\sin k_j - \lambda_\alpha - iU/4}, \\
&&\Pi_{j = 1}^N \frac{\lambda_\alpha - \sin k_j + iU/4}{\lambda_\alpha - \sin k_j - iU/4} = 
-\Pi_{\beta= 1}^{N_\downarrow} \frac{\lambda_\alpha - \lambda_\beta + iU/2}{\lambda_\alpha - \lambda_\beta - iU/2}, 
\end{eqnarray}
where $L/N$ is the total number of sites/fermions and $N_{\uparrow/\downarrow}$ is the number of spin up/down fermions $(N_\downarrow \le N/2)$. The pseudo-momentum $\{k_j\}$ and spin rapidities $\{\lambda_\alpha \}$ are variables to be solved and are related to the total energy by $E = -2t \sum_{j = 1}^N \cos k_j$ and the total momentum by $p = \sum_{j = 1}^N k_j$. For simplicity, here we just consider the energy spectrum of holon excitations in the charge sector \cite{fhba2}. Results of spinon excitations in the spin sector can also be obtained similarly.

In Fig. \ref{fig:FH_L18N6_DL_pred}(a) and (b), we show the predicted energy spectrum, $(p_i,E_i)$, obtained by RSNN approach at the intermediate ($\lambda\equiv U/(t+U)=0.53$ or $U/t \sim 1$) and strong interaction strength ($\lambda=0.89$ or $U/t \sim 9$) respectively. The training regime is in the weakly interacting regime $R_\textrm{train}=(0.1,0.35)$. The input features are $16\times 16$ random sampling matrices, obtained from the original system Hamiltonian (see Fig.\ref{fig:method} and Sec. \ref{sec:concept}), and the output label is the whole energy spectrum, $(p_i,E_i)$. In the training regime, we take $N_{train}=2000$ values of $\lambda$s and generate $N_{s}=100$ sampling matrices for each of them. Compared to the exact solution by BA, the predicted results of RSNN are pretty good even in the strongly interacting regime. In (c) we show the associated density of states (DOS), which reflects the interaction-broadened band widths.

In Fig.~\ref{fig:FH_L18N6_DL_pred}(d), we show the obtained accuracy and its uncertainty (obtained by averaging five independent calculations) for the whole test regime. We could find that the accuracy of the whole energy spectrum could be as high as 99\% in the intermediate interaction regime ($\lambda\sim 0.5$ or $U/t\sim 1$), while it decreases gradually in the strongly interacting regime ($\lambda> 0.8$ or $U/t>4$) with a larger uncertainty at the same time. However, we note that it is still impressive that the accuracy can be still larger than 95\% even for $\lambda=0.923$ ($U/t \sim 12$), showing that RSNN could be a very promising tool to estimate physical quantities in a strongly correlated system even it is trained in the weakly interacting regime.

\section{Quantum Critical Exponents of 1D $XXZ$ Model}
\label{sec:XXZ}

The quantum phase transition we discussed in the 1D IMTF is a first order transition, where the magnetization changes discontinuously in the thermodynamic limit ($N\to\infty$). It is therefore instructive to see if RSNN could be also applied to the study of the second order phase transition, where the order parameter changes continuously in the thermodynamic limit and hence scaling exponents could be identified near the quantum critical point (QCP) \cite{subir11}.

One of the most important examples is the superfluid to Mott Insulator transition for strongly interacting bosonic atoms loaded in an optical lattice \cite{sfmi}. Here we consider a simpler case of hard-core bosons with a finite inter-site repulsion to compete with the kinetic energy, leading to the so-called Bose $t-V$ model:
\begin{eqnarray}
\label{eq:t-V}
\hat{\cal H}_{t-V}&=&\sum_{i=1}^N
\left[-t\left(\hat{b}_{i}^\dagger \hat{b}_{i+1}+h.c.\right)+V\hat{n}_i\hat{n}_{i+1}-\mu \hat{n}_{i}\right],
\end{eqnarray}
where $\hat{b}_i$ and $\hat{n}_i=\hat{b}_i^\dagger\hat{b}_i$ are bosonic field operator and number operator respectively; $t$ and $V$ are the tunnelling and interaction between the neasrest neighboring sites; $\mu$ is the chemical potential. It is easy to see that the system prefers to be superfluid if $V$ is small, and can become a solid phase at half-filling when $V$ is repulsive and large. Quantum phase diagrams of such superfluid to solid transition has been studied by quantum Monte Carlo methods in 1D and 2D systems \cite{qmc1}.

Since the number of particle per site is either 0 or 1 in the hard-core limit, it is easy to connect such a $t-V$ model of hard-core bosons to a spin $1/2$ system. More precisely, in the dilute limit, one could use Holstein and Primakoff transformation \cite{hpt} to map the above $t-V$ model into spin-$1/2$ $XXZ$ model with a transverse field. In order to simplify the calculation in the rest of this paper, however, we will concentrate on the quantum phase transition of the 1D $XXZ$ model itself at zero field, which has the following system Hamiltonian: 
\begin{eqnarray}
\hat{\cal H}_{XXZ} = -\frac{J}{2}\sum_{j = 1}^{N}  (\hat{\sigma}_j^x \hat{\sigma}_{j+1}^x +\hat{\sigma}_j^y \hat{\sigma}_{j+1}^y +\lambda \hat{\sigma}_j^z \hat{\sigma}_{j+1}^z),
\label{eq:XXZ}
\end{eqnarray}
where $J$ is the in-plane spin coupling and $\lambda$ is the $z$-direction spin coupling.
It is well-known that there are three different phases for 1D XXZ model in the thermodynamic limit: anti-ferromagnetism (AFM, gapped) for $\lambda < -1$, paramagnetism (PM, gapless) for $-1 < \lambda < 1$, and ferromagnetism (FM, gapped) for $\lambda > 1$. The superfluid to solid transition of the $t-V$ model of hard-core bosons corresponds to the AFM-PM transition at $\lambda=-1$, which we will study closely by RSNN here.
We note that the spin $1/2$ 1D XXZ model in Eq. (\ref{eq:XXZ}) could be also exactly solved by Bethe-ansatz method \cite{cnyang1, cnyang2, cnyang3}. 

\begin{figure}
\centering
\includegraphics[scale=0.24]{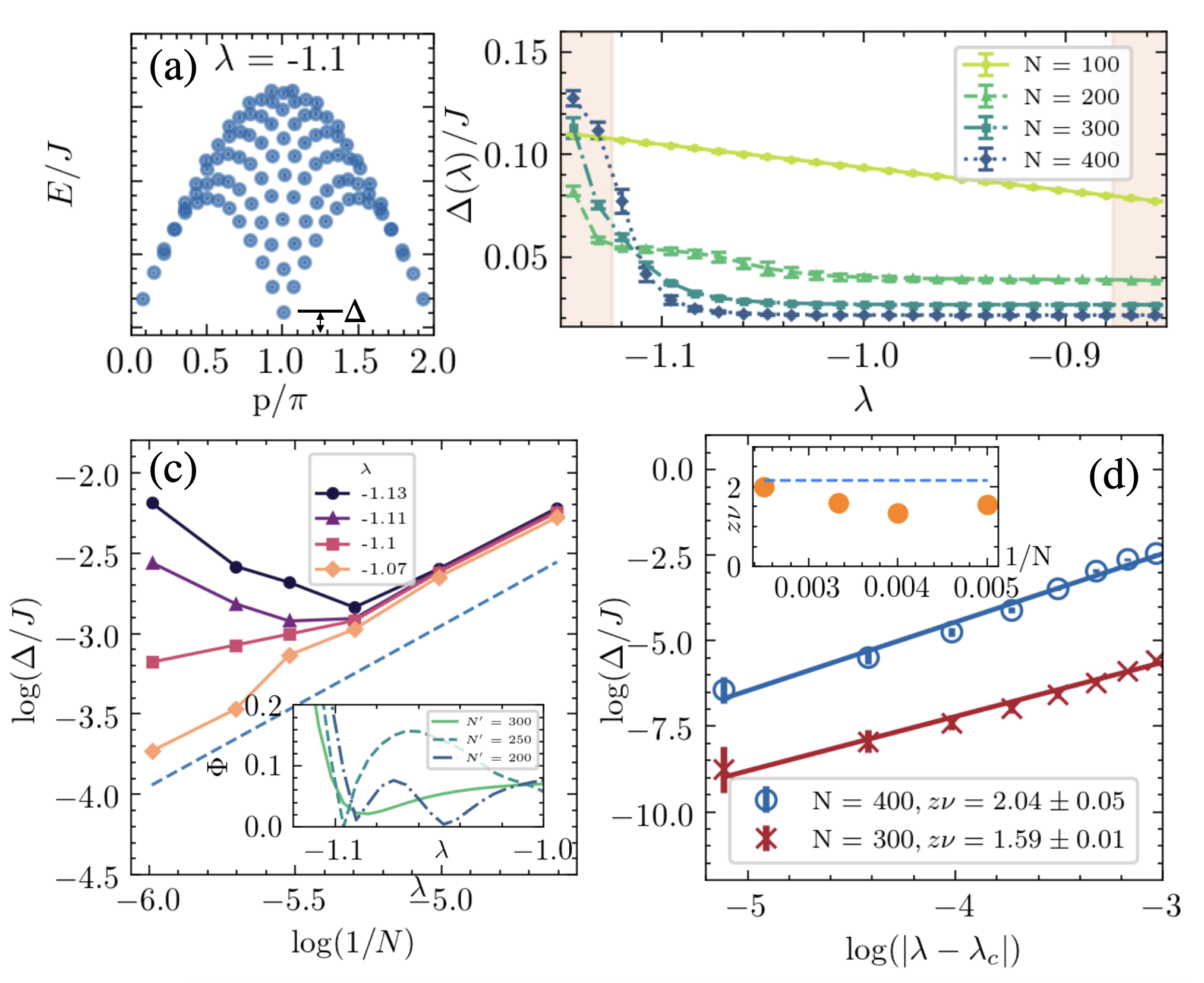}
\caption{
(a) Spinon excitation spectrum of 1D XXZ model for $N = 28$ and $\lambda = -1.1$. The spin excitation gap, $\Delta$, is defined by the excitation energy at $p=\pi$. (b) The spin excitation gap, $\Delta$, predicted by RSNN for several system sizes up to $N = 400$ with the corresponding uncertainties. The training/test regime is indicated by the colored/white background. 
(c) RSNN predicted $\Delta$ as a function of $1/N$ (in a log-log plot) for different values of $\lambda$s. The predicted phase transition point, $\lambda_c^\textrm{RSNN}=-1.07 \pm 0.02$, is defined as the fixed point of Phenomenological Renormalization Group (see the text). The dashed line is the slope given by $\lambda_c^\textrm{RSNN}$. The inset shows the obtained $\Phi(\lambda)$ (see the text) for several system sizes. The position of its minimum value gives an estimate of critical quantum phase transition point. (d) Predicted $\Delta$ as a function of $|\lambda-\lambda_c^\textrm{RSNN}|$ (in a log-log plot) for $M=300$ and 400. The solid lines are the fitting results for different values of dynamical exponent. Inset shows how such critical exponent ($z\nu$) changes as a function of the system size. $z\nu=2.16$ in the thermodynamic limit, as indicated by the horizontal dashed line.
}
\label{fig:xxz_N28_spinon_excitation}
\end{figure}

In order to investigate the quantum phase transition point near $\lambda=-1$,
we use $\lambda \in (-1.15, -1.125)\cup(-0.875, -0.85)$ as the training regime with $N_\textrm{train}=500$ for the training data inside. For each $\lambda$, we generate $N_s=100$ random sampling matrices (with the dimension $M=10$) as the input features. In Fig.~\ref{fig:xxz_N28_spinon_excitation}(a), we show the predicted spectrum for $\lambda = -1.1$ (in the test regime) for $N=28$. The obtained spectrum agrees with the BA results (not shown) very well.  The lowest energy excitation occurs at $ p = \pi$ as expected. The average accuracy of the whole energy spectrum is $98.01 \pm 1.29 \%$ in the whole test regime, showing a pretty good prediction even near the phase transition point, $\lambda=-1$.

In Fig. \ref{fig:xxz_N28_spinon_excitation}(b) we show the calculated spinon excitatin gap, $\Delta$, as a function of $\lambda$ for various system sizes, $N$. We could find that the gap becomes almost vanished for $\lambda>-1$ as $N> 300$. However, as the system size increases, the matrix elements in the sampling Hamiltonians cover less and less fraction of the original Hilbert space and therefore the prediction accuracy also decreases down to $90.49 \pm 1.09 \%$ for $N = 300$.

Different from the first order phase transition of 1D IMTF, the order parameter (here the spinon excitaiton gap) should decrease to zero continuously at the QPT point ($\lambda_c$) in the thermodynamic limit, but it always has a finite value for a finite $N$. In order to determine the QCP from the finite size scaling, here we use the phenomenological renormalization group (PRG) method \cite{prg, prg2}: Using the fact that the excitation gap must scale with $N$ linearly at the QCP in a 1D system, i.e. $\Delta(N,\lambda_c)\propto N^{-1}$, it is reasonable to expect that, for any two large system sizes, $N\neq N'\gg 1$,  $\Phi(N,N',\lambda)\equiv|1-N'\Delta(N',\lambda)/N\Delta(N,\lambda)|$ has a minimum at $\lambda_\textrm{min}(N,N')$. This minimum value of $\Phi$ should reach zero and $\lambda_\textrm{min}(N,N')\to \lambda_c$ when $N,N'\to\infty$ at the same time. As a result, after considering the possible uncertainty of finite size calculation, we define $\lambda_c^\textrm{RBNN}(N)\equiv Ave_{N'}[\lambda_\textrm{min}(N,N')]$, where $Ave_{N'}[\cdots]$ is the average of different system sizes, $N'$, by keeping another one ($N$) fixed.

In Fig. \ref{fig:xxz_N28_spinon_excitation}(c) we show the gap as a function of $1/N$ in a log-log plot with different values of $\lambda$s near the quantum critical point (QCP), $\lambda_c=-1$. We could see that the curves approaches linear when $\lambda$ is increased from below $\lambda_c$ as expected. In the inset, we show the calculated function, $\Phi(N,N',\lambda)$, for $N=400$ and $N'=200$, 250, and 300 respectively. We could see that there are two local minimum values near -1.1 and -1.05. After averaging the position of the minimum values, we obtain the estimated QCP at $\lambda_c^\textrm{RSNN}=-1.07 \pm 0.02$, pretty close to the value $\lambda_c^\textrm{BA}=-1.06$ obtained by finite size Bethe-ansatz method. However, we have to emphasize that for the results calculated by BA in the same regime (not shown here), $\Phi(N,N',\lambda)$ is a pretty flat function with only one shallow minimum, different from the results predicted by RSNN. Therefore, what we could say is that the results predicted by RSNN here is just an estimate of the QCP, based on the data trained outside the critical regime. 
  
Finally, using the obtained QCP value, $\lambda_c^\textrm{RSNN}=-1.07$, we could further calculate the critical exponent $z \nu$, which is defined by how the gap function vanishes~\cite{subir11} near the QCP: $\Delta \sim | \lambda - \lambda_c |^{z \nu}$ for $\lambda<\lambda_c$ in the thermodynamic limit (note that $\Delta=0$ for $\lambda>\lambda_c$). In Fig.~\ref{fig:xxz_N28_spinon_excitation}(d), we show that such a nontrivial scaling exponent could be obtained to be $z\nu=2.04 \pm 0.05$ by RSNN, and it is close (within 5\% uncertainty) to the numerical value, $z \nu = 2.16$, obtained by the Bethe-ansatz in thermodynamics limit \cite{bm} (see Appendix \ref{sec: nu}. Note that the value of $\nu$ is known to be one for 1D XXZ model \cite{nu1, nu2}). Again, we find RSNN could provide a reasonably good estimate of the critical exponent even if using the data outside the critical regime.  

\section{Summary}
\label{sec:summary}

Motivated by the pattern recognition method in computer vision, We propose a new approach to predict the physical quantities of a general many-body system by randomly sampling the whole system Hamiltonian through a self-supervised learning process. The training data could be obtained by perturbation theory or other existing numerical methods in the weakly interacting regime (or any certain parameter regimes). We have systematically investigate its applicability in several 1D exactly solvable models, and demonstrate how it could provide pretty good prediction results of the ground state energy, the momentum-energy spectrum, magnetization (or other order parameters), as well as the quantum phase transition point and the associated critical exponents. One of the most important advantages of RSNN is that one just needs to train the model one time in the training regime and then gets an arbitrary amount of data immediately in the test regime, even in the strongly correlated regime or near the quantum phase transition point. Combination of RSNN with other numerical methods may provide a very effective approach to explore quantum many-body problems. 

\begin{center}
    \textbf{CODE AVAILABILITY}
\end{center}

We provide Github code (\url{https://github.com/CYLphysics/RSNN_TFIM1D}) for the application of RSNN in 1D Ising Model with a Transverse Field.

\acknowledgments{
We thank Ming-Chiang Chung, Pochung Chen, Chung-Yu Mou, and Chung-Hou Chung for fruitful discussions. This work is supported by the Ministry of Science and Technology grant (MOST 107-2112-M-007-019-MY3) and by the Higher Education Sprout Project funded by the Ministry of Science and Technology and the Ministry of Education in Taiwan. We thank the National Center for Theoretical Sciences for providing full support. }

\nocite{*}

\bibliography{CNNMBS}

\appendix

\section{Model Parameters of RSNN.}
\label{sec:model}

In the construction of RSNN, as introduced in Sec. \ref{sec:concept}, we have two types of hyper-parameters: one is about the process for random sampling and the other is about the structure of CNN structure, which accepts the input of random sampling matrices and outputs the expected physical quantities through neural networks (see also, Fig. \ref{fig:method}). Since we have mentioned the parameters used for random sampling in the text for each model (say, matrix size, $M$, number of sampling matrices, $N_S$, number of training data, $N_\textrm{train}$, and training regime, $R_\textrm{train}$ etc.), here we provide further information about the hyper-parameters used for the second part in the CNN structure.

As for the final output layer, we use a loss function to constrain the output to be our desired values for a self-supervised learning process. For example, if the output is to simulate the three lowest exactly known eigenenergies of the 1D IMTF system as shown in Sec. \ref{sec:IMTF}), the loss function we used is designed as following:
\begin{eqnarray}
\textit{Loss} = \overline{|\textit{Pred} - \vec{E}^\textrm{ED}|} + \beta \sum_i (W_i^2+b_i^2),
\label{eq:loss}
\end{eqnarray}
Where $\textit{Pred}$ is the predicted results of the neural network for each run and $\vec{E}^\textrm{ED}\equiv\left[E^\textrm{ED}_0,E^\textrm{ED}_1,E^\textrm{ED}_2\right]$ are the lowest three eigenstate energies provided by exact diagonalization (or results known before) respectively. The first term is taken as the batch average, and second term is to constrain the magnitudes of weighting ($W_i$) and bias ($b_i$) of all neurons (with index $i$) from over-fitting. $\beta>0$ is an empirical parameter. The optimization process is done by Adam method \cite{adam} with the corresponding model parameters shown in Tab.\ref{MP}.  
\begin{table}[ht]
\begin{center}
\centering
\begin{tabular}
{ |p{3.5cm}|p{1.0cm}|p{1.0cm}|p{2cm}|}
 \hline
\diagbox[width=8em]{Case}{Parameter} & $L_{\text{conv}}$ & $L_{\text{fc}}$ & $N_{\text{neuron}}$  \\
 \hline
$\text{IMTF}(eigenenergies)$ & 2 & 2 & [1250,10] \\

$\text{IMTF}(magnetization) $ & 2  & 2 & [625,100] \\

\text{FHM}(holon spectrum)  & 2 & 6 & [625,102,102, \\
    &   &   &  102,102,102]  \\

$\text{XXZ}(spinon spectrum)$  & 2 & 6 & [625,102,102, \\
                       &   &   &  102,102,102]  \\
             
$\text{XXZ}(spinon gap)$ & 2 & 3 & [625,40,20] \\

 \hline

\end{tabular}
\label{MP}
\end{center}
\caption{Hyper-parameters used for the CNN part of RSNN  in the examples of this paper (see the text). $L_{\text{conv}}$ refers to the number of convolutional layers , $L_{\text{fc}}$ refers to the number of fully connected layers and $N_{\text{neurons}}$ is the number of neurons for each layer as shown in the list.}
\end{table}

\section{Exponent $z \nu$ by Bethe-ansatz.}
\label{sec: nu}

In order to extract the critical exponent of 1D $XXZ$ model, we calculate it from the Bethe-ansatz solution in the thermodynamic limit ($L \rightarrow \infty$). For $\lambda < -1$, the dispersion relation in thermodynamics limit can be expressed in the integral form as following \cite{bm}:
\begin{eqnarray}
\label{eq:dipsersion}
E(P) = \frac{2K(m) \sinh \phi(\lambda)}{\pi} \sqrt{1 - m \sin^2 P},
\end{eqnarray}
where $E(P)$ is the energy dispersion of momentum $P$, 
$\lambda = - \cosh \phi$, $\phi \equiv \pi K'(m)/K(m)$, parameter $m = k^2$ with $k$ the elliptic moduluss, and $K(m)$ is the complete elliptic integral of first kind:
\begin{eqnarray}
K(m) = \int_0^{\pi/2} \frac{d\theta}{\sqrt{1 - m \sin^2 \theta}},
\end{eqnarray}
With a fixed $\lambda$, we can obtain the value of $m$ by solving the differential equation $\cosh^{-1}(-\lambda) = \pi K'(m)/K(m)$, which evaluate
the dispersion function eq.(\ref{eq:dipsersion}). Such function exists a lowest energy gap at $P = \pi/2$, thus the gap function $\Delta (\lambda)$ is defined as:
\begin{eqnarray}
 \Delta(\lambda) \equiv E(\pi/2) = \frac{2 K(m) \sinh \phi(\lambda)}{\pi} \sqrt{1-m},
\end{eqnarray}
 which has a fitting parameter (exponent) $z \nu = 2.16$ for the form $|\lambda - \lambda_c|^{z \nu}$ with $\lambda_c = -1$.


\end{document}